\documentclass[aps,prb,twocolumn,showpacs,amsmath,amssymb]{revtex4}

\usepackage{graphicx}
\usepackage{dcolumn}
\usepackage{bm}
\begin{document}

\title{Dynamics of the Magnetic Flux Trapped in Fractal Clusters \\of a Normal Phase
in Percolative Superconductors}

\author{Yuriy I. Kuzmin}
\email{yurk@mail.ioffe.ru}

\affiliation{Ioffe Physical Technical Institute of the Russian
Academy of Sciences, 26 Polytechnicheskaya Street, Saint
Petersburg 194021 Russia}

\date{\today}

\begin{abstract}
The effect of the fractal clusters of a normal phase, which act as
pinning centers, on the dynamics of magnetic flux in percolative
type-II superconductor is considered. The main features of these
clusters are studied in detail: the cluster statistics is
analyzed; the fractal dimension of their boundary is estimated;
the distribution of critical currents is obtained, and its
peculiarities are explored. It is found that there is the range of
fractal dimension where this distribution has anomalous
statistical properties, specifically, its dispersion becomes
infinite. It is examined how the finite resolution capacity of the
cluster geometric size measurement affects the estimated value of
fractal dimension. The effect of fractal properties of the normal
phase clusters on the electric field arisen from magnetic flux
motion is investigated for the cluster area distribution of
different kinds. The voltage-current characteristics of fractal
superconducting structures in the resistive state are obtained for
an arbitrary fractal dimension. It is revealed that the fractality
of the boundaries of the normal phase clusters intensifies the
magnetic flux trapping and thereby raises the critical current of
a superconductor.
\end{abstract}

\pacs{74.81.-g; 74.25.Qt; 74.25.Sv, 74.25.Fy; 74.81.Bd}

\maketitle

\section{Introduction}

An essential feature of the clusters of columnar defects in
superconductors consists in their capability to trap a magnetic
flux. \cite{tonomura,lidmar,dam} Holding in place the vortices
driven by the Lorentz force, such clusters can act as effective
pinning centers. \cite{higuchi,mezzetti} This feature can be used
in making new composite superconducting materials of enhanced
critical current value. \cite{beasley,krusin,tpl} The
characteristics of the normal phase clusters exert an appreciable
effect on vortex dynamics in superconductors, especially when the
clusters have fractal boundaries.
\cite{olson,surdeanu,prester,pla,prb} In the present paper the
magnetic and transport properties of composite superconductors
with fractal clusters of a normal phase will be reviewed as well
as the phenomena limiting the current-carrying capability of such
superconductors will be considered.

Let us consider a superconductor containing columnar inclusions of
a normal phase, which are out of contact one with another. These
inclusions may be formed by the fragments of different chemical
composition, as well as by the domains of reduced superconducting
order parameter. The similar columnar defects can readily be
created during the film growth process at the sites of defects on
the boundary with the substrate. When such a superconducting
structure is cooled below the critical temperature in the magnetic
field oriented along the direction of the longest size of these
inclusions, the magnetic flux will be frozen in the normal phase
clusters. By the cluster we mean a set of the columnar defects,
which are united by the common trapped flux and are surrounded by
the superconducting phase. Even after the external field has been
turned off, the flux trapped in these clusters is kept unchanged
due to the currents that are steadily circulating around them
through the superconducting loops. The distribution of the trapped
magnetic flux resulting from such a magnetization in the
field-cooling regime will be two-dimensional. So, instead of
dealing with an extended object, which indeed the normal phase
cluster is, we will consider its cross-section by the plane
carrying a transport current. We will examine the geometric
properties of the normal phase clusters in the planar section
only, where the boundaries of the clusters are statistically
self-similar.

The paper is organized as follows: The setting of the problem is
described as well as the relationship between magnetic flux
trapping, pinning properties, and geometric characteristics of the
normal phase clusters in superconductor is discussed in Sec. 2. A
geometric probability analysis of the weak links distribution over
the cluster perimeter in terms of path integrals is made in Sec.
3. The concept of fractals and the basic notions in this field are
reviewed briefly in Sec. 4. The statistical properties of normal
phase clusters in YBCO superconducting films are studied, and the
fractal dimension of their boundaries is estimated in Sec. 5. The
voltage-current characteristics of superconductors with fractal
clusters of a normal phase are obtained, and the peculiarities of
the resistive state of the fractal superconducting structures are
considered in Sec. 6.

\section{Magnetic Flux Trapping in Percolative Superconductors}

\begin{figure}
\includegraphics{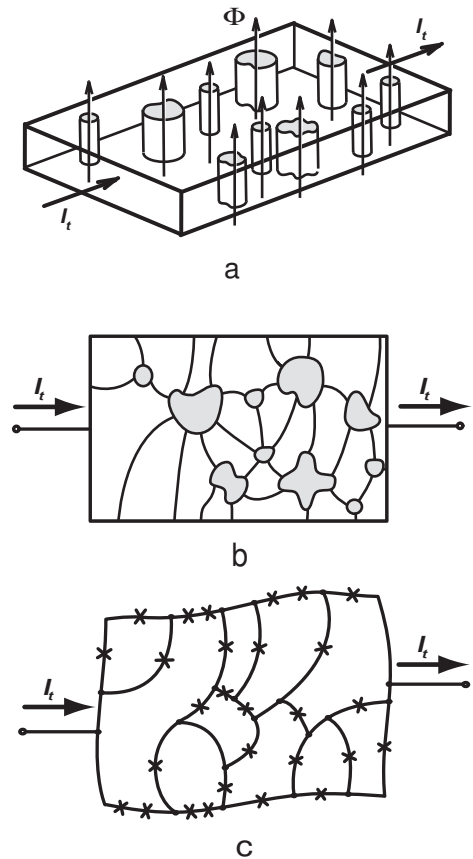}
\caption{\label{fig1} Schematic drawing of superconductor with a
magnetic flux trapped in the clusters of columnar defects (a), its
percolation representation in the cross-section by the plane
carrying a transport current (b), and the equivalent
weak-link-network circuit (c). The sections of normal phase
clusters are shaded. In picture (b) weak links are shown by the
curves joining together the normal phase clusters, and, in picture
(c), by the crosses on the superconducting loops that surround the
normal phase inclusions.}
\end{figure}

The situation is sketched out in Fig.~\ref{fig1}(a). The composite
superconductor containing clusters of a normal phase represents a
percolation system, where both the electric percolation of the
supercurrent and the percolation of a magnetic flux may happen. If
the part of the film surface covered by the normal phase exceeds
the percolation threshold (which is equal to 50 percents for
two-dimensional percolation \cite{stauffer}), the magnetic flux
could be free to move in the transversal direction. There is no
pinning at all in this situation, so it is out of our interest.
Let us suppose that the opposite case is realized when the
relative portion of superconducting phase exceeds the percolation
threshold (see Fig.~\ref{fig1}(b)), so there is a superconducting
percolation cluster in the plane of the film where a transport
current can flow. Such a structure provides for effective pinning
and thereby raises the critical current, because the magnetic flux
is locked in finite clusters of a normal phase, and so the
vortices cannot leave them without crossing the surrounding
superconducting space. Let us denote the total magnetic flux
trapped in the superconductor after the field-cooling
magnetization by $\Phi$. If the transport current is passed
through the sample, the trapped flux remains unchanged as long as
the vortices are still held in the normal phase clusters. The
larger clusters have to provide for a weaker pinning than smaller
ones, because the larger the cluster size, the more weak sites,
through which the vortices can pass under the action of the
Lorentz force, can be located along its perimeter. When the
current is increased, the magnetic flux will break away first from
the clusters of smaller pinning force, and therefore, of larger
size. Magnetic flux trapped into a single cluster is proportional
to its area $A$. Therefore, the decrease in the total trapped flux
$\Delta \Phi $ is proportional to the number of the clusters of
area larger than a given magnitude $A$. So it can be expressed
with the cumulative probability function $W\left( A\right) =\Pr
\left\{ \forall A_{j}<A\right\} $, which is equal to the
probability to find the cluster of area $A_{j}$ smaller than a
preset value of $A$:
\begin{equation}
\frac{\Delta \Phi }{\Phi }=1-W\left( A\right)  \label{proba1}
\end{equation}

The left hand side of this formula is equal to the relative
decrease in the total trapped flux caused by the transport current
of the same amplitude as the depinning current of the cluster of
area $A$, and the right hand side gives the probability to find
the cluster of area greater than $A$ in the whole population.

When the magnetic flux starts to break away from the normal phase
clusters, the vortices must cross the surrounding superconducting
space, and they will first do that through the weak links, which
connect the normal phase clusters between themselves (see
Fig.~\ref{fig1}(b) where weak links are shown by the curves
connecting the shaded regions, which designate the normal phase
clusters). In Fig.~\ref{fig1}(c) the equivalent weak-link-network
circuit representation of such a system is shown. The normal phase
clusters are presented by the cells of the net interlaced by the
superconducting loops. Each of the loops contains weak links
(shown by the crosses), which join the adjacent cells and so
enable the vortices to pass from one cluster to another.

Weak links form readily in high-temperature superconductors (HTS)
characterized by an extremely short coherence length.
\cite{blatter,scalapino,kerchner} Various structural defects,
which would simply cause some additional scattering at long
coherence length, give rise to the weak links in HTS. On a
mesoscopic scale twin boundaries are mainly responsible for weak
link existence. \cite{pastoriza,kupfer,maggio,liu} Twins form
especially readily in YBCO superconductors inasmuch as their unit
cell is only close to the orthorhombic one. The twins can be
spaced up to several nanometers apart, so even single crystal may
have the fine substructure caused by twins. Magnetic and transport
properties of HTS depend strongly on the orientation of twin
planes with respect to the applied magnetic field.
\cite{oussena1,oussena2} The flux can easily move along the weak
links formed by twins. \cite{duran1,duran2,welp,wijn} At last, on
a macroscopic scale there are manifold structural defects which
can form weak links: that may be grain or crystallite boundaries
as well as barriers arising from the secondary degrading the
non-stoichiometric crystal into the domains with a high and low
content of oxygen. \cite{kerchner,hanisch,rykov,schuster,kilic}
Moreover, a magnetic field further reduces a coherence length,
thus resulting in more easy weak links formation. \cite{sonier} In
conventional low-temperature superconductors the weak links can be
formed due to the proximity effect in sites of minimum distance
between the next normal phase clusters.

\begin{figure}
\includegraphics{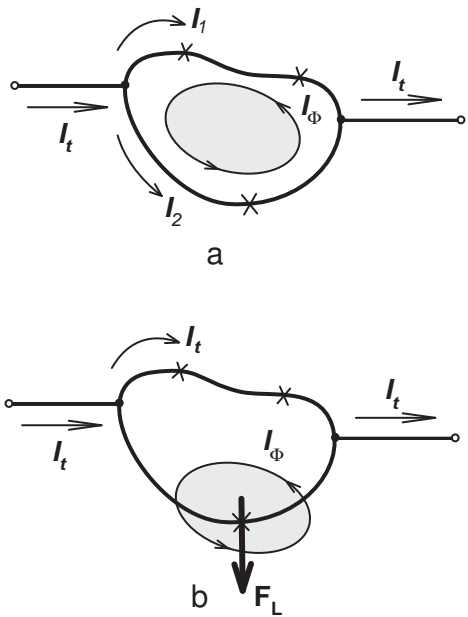}
\caption{\label{fig2} A simple illustration how the magnetic flux
exits from a normal phase cluster through the weak link. Here
$I_{t}=I_{1}+I_{2}$ is the transport current; $I_{\Phi }$ is the
current related to the trapped magnetic flux $\Phi $; ${\bf
F}_{L}$ is the Lorentz force. (a) - magnetic flux remains to be
locked in the cluster as long as $I_{2}+I_{\Phi }<I_{c}$,
where $I_{c}$ is the critical current of the weak link; (b) - as soon as $%
I_{2}+I_{\Phi }\ge I_{c}$ the Lorentz force expels the flux
through the right weak link which has become resistive.}
\end{figure}

As soon as the transport current is turned on, this one is added
to all the persistent currents, which maintain the magnetic flux
to be trapped. The
situation is sketched in Fig.~\ref{fig2}. Each of these currents (like $%
I_{\Phi }$) is circulating through the superconducting loop around
the normal phase cluster wherein the corresponding portion of the
magnetic flux is locked (Fig.~\ref{fig2}(a)). The loop contains
weak links that join the adjacent normal phase clusters
transversely to the path of the current. As the transport current
is increased, there will come a point when the overall current
flowing through the weak link will exceed the critical value, so
this link will turn into a resistive state. As this takes place,
the space distribution of the currents throughout the
superconducting cluster is changed in such a way that the
resistive subcircuit will be shunted by the superconducting paths
where weak links are not damaged yet. Magnetic field created by
this re-distributed transport current acts via the Lorentz force
on the current circulating around the normal phase cluster. As a
result, the magnetic flux trapped therein will be forced out
through the resistive weak link, which has become permeable to the
vortices (Fig.~\ref{fig2}(b)). So the cluster will not contain the
magnetic flux any more.

Thus, whatever the microscopic nature of weak links may be, they
form the channels for vortex transport. It appears that according
to their configuration each normal phase cluster has its own value
of the critical current, which contributes to the overall
statistical distribution. By the critical current of the cluster
we mean the current of depinning, that is to say, such a current
at which the magnetic flux ceases to be held inside the cluster of
a normal phase. When a transport current $I$ is gradually
increased, the vortices will break away first from clusters of
small pinning force (weaker than the Lorentz force created by the
transport current), and therefore, of small critical current. Thus
the decrease in the trapped magnetic flux $\Delta \Phi $ is
proportional to the number of all the normal phase clusters of
critical currents less than a preset value. Therefore, the
relative change in the trapped flux can be represented by the
cumulative probability function $F\left( I\right) =\Pr \left\{
\forall I_{j}<I\right\} $ for the distribution of the critical
currents of clusters:
\begin{equation}
\frac{\Delta \Phi }{\Phi }=F\left( I\right)  \label{probi2}
\end{equation}

It is obvious that the critical current distribution $F=F\left(
I\right) $ is related to the cluster area distribution $W=W\left(
A\right) $, because the cluster of a larger size has more weak
links over its boundary with the surrounding superconducting
space, and thus the smaller current of depinning.

\section{Geometric Probability Analysis of the Distribution of Weak Links
over the Cluster Perimeter}

In order to find out the relationship
between the distribution of the critical currents of the clusters
and the distribution of their areas, the distribution of entry
points into weak links over the perimeter of a normal phase
cluster should be analyzed. The problem of exit of a vortex from a
normal cluster represents the two-dimensional analogue of a
problem of a random walk particle reaching a border.
\cite{spitzer,gardiner} At the same time, unlike the classic
problem on the distribution of the exit points, here the boundary
of the area is not absorbing all over. There are only discrete
absorption points, which are located just at the sites where weak
links are going out on the cluster boundary. In other words, these
points are the points of the entry of vortices into weak links, or
simply, entry points. For simplicity, we will assume that after
the vortex reaches the entry point, it passes all the way between
two adjacent normal phase clusters without being trapped inside
the weak link itself. Here the magnetic flux is transferred by
Josephson vortices. The Josephson penetration depth is large
enough in the considered materials, so the size of the region,
where the vortex is localized, much exceeds the characteristic
length of all possible structural defects that can occur along the
transport channel. Thus the probability that such a vortex, driven
by the Lorentz force, will be trapped in passing through a weak
link is very small. This assumption agrees well with the results
of research on the magnetic flux motion along weak links
\cite{wijn,dorog,fangohr} including twins.
\cite{duran1,duran2,welp,turch,welp2} At the same time, it allows
us to highlight the role played by the cluster boundary in the
magnetic flux dynamics.

Let us consider the distribution of entry points over perimeter of
a normal phase cluster. Generally, this distribution varies from
one cluster to another, so that each normal phase cluster has the
entry point distribution function $\psi \left( l\right) $ of its
own, which belongs to some function class $\Omega $. Here $l$ is
the co-ordinate measured along the cluster perimeter, so $l\in
\left( 0,P\right) $. In this context the functions of class
$\Omega $ are random elements of the statistical distribution. The
probability distribution of functions $\psi \left( l\right) $ over
all the clusters can be characterized by the functional $\Pr
\left\{ \psi \left( l\right) \right\} $, which is equal to the
probability of finding a given function $\psi \left( l\right) $.

In the most general way the geometric probability analysis of the
entry point into weak link distribution can be carried out by
means of path integral technique. \cite{feynman} The probability
that the function $\psi \left( l\right) $ may be found within the
class $\Omega $ can be expressed by the path integral
\[
\Pr \left\{ \psi \left( l\right) \right| \Omega
\}=\int\limits_{(\Omega )}D\psi \left( l\right) \Pr \left\{ \psi
\left( l\right) \right\}
\]

Therefore, the most probable function of entry point distribution
is the
mean over all functions of class $\Omega $%
\begin{equation}
\Psi \left( l\right) \equiv \overline{\psi \left( l\right) }%
=\int\limits_{(\Omega )}D\psi \left( l\right) \psi \left( l\right)
\Pr \left\{ \psi \left( l\right) \right\}  \label{pathprob4}
\end{equation}
The path integral Fourier transform on the probability functional
$\Pr \left\{ \psi \left( l\right) \right\} $ represents the
characteristic functional
\begin{equation}
H\left[ k(l)\right] =\frac{\int\limits_{\left( \Omega \right)
}{\cal D}\psi \left( l\right) \,\exp \left( i\oint dl\,k\left(
l\right) \psi \left(
l\right) \right) \Pr \left\{ \psi \left( l\right) \right\} }{%
\int\limits_{\left( \Omega \right) }{\cal D}\psi \left( l\right)
\,\Pr \left\{ \psi \left( l\right) \right\} }  \label{char5}
\end{equation}
where $k=k\left( l\right) $ are the functions of a reciprocal
function set, and integration in the kernel $\exp \left( i\oint
dl\,k\left( l\right) \psi \left( l\right) \right) $ is carried out
over the cluster perimeter.

The characteristic functional is the path integration analog for
the usual moment-generating function. The probability functional
$\Pr \left\{ \psi \left( l\right) \right\} $ can be written as an
inverse path integral Fourier transform on the characteristic
functional
\[
\Pr \left\{ \psi \left( l\right) \right\} =\int Dk\left( l\right)
\exp \left( -i\oint dl\,k\left( l\right) \psi \left( l\right)
\right) H\left[ k\left( l\right) \right]
\]
where the path integration is carried out on the reciprocal
function space.

In the simplest case, when all the clusters are of equal entry
point distribution, which coincides with the most probable one of
Eq.~(\ref {pathprob4}), the probability functional $\Pr \left\{
\psi \left( l\right) \right\} $ is zero for all $\psi \left(
l\right) $ that differ from $\Psi \left( l\right) $, whereas $\Pr
\left\{ \Psi \left( l\right) \right\} =1$. As this takes place the
characteristic functional of Eq.~(\ref{char5}) becomes
\begin{equation}
H\left[ k\left( l\right) \right] =\exp \left( i\oint dlk\left(
l\right) \Psi \left( l\right) \right)  \label{charfin6}
\end{equation}

If all entry points had fixed co-ordinates $l_{j}$ instead of the
random ones, their distribution would be $\psi \left( l\right)
=\beta
\mathop{\textstyle\sum}%
\nolimits_{j=1}^{N}\delta \left( l-l_{j}\right) $ , where $N$ is
the number of entry points along the cluster perimeter, $\delta
\left( l\right) $ is Dirac delta function. The constant $\beta $
is being chosen to normalize the distribution function $\psi
\left( l\right) $ to unity, so that $\beta N=1$.

Now suppose that all the points of entries into weak links are
randomly distributed with uniform probability over the cluster
perimeter, so the probability to find any $j$-th point within some
interval $dl_{j}$ is proportional to its length. In that case the
characteristic functional of Eq.~(\ref{char5}) takes the form
\begin{eqnarray}
H\left[ k\left( l\right) \right] &=&\frac{\oint
\mathop{\textstyle\prod}%
\limits_{j=1}^{N}dl_{j}\,\exp \left( i\beta
\sum\limits_{j=1}^{N}\oint dl\,k\left( l\right) \delta \left(
l-l_{j}\right) \right) }{\oint
\prod\limits_{j=1}^{N}dl_{j}}=  \nonumber \\
&=&\frac{1}{P^{N}}\oint
\mathop{\textstyle\prod}%
\limits_{j=1}^{N}dl_{j}\,\exp \left( i\beta
\sum\limits_{j=1}^{N}k\left(
l_{j}\right) \right) = \nonumber \\
&=&\frac{1}{P^{N}}%
\mathop{\textstyle\prod}%
\limits_{j=1}^{N}%
\displaystyle\oint %
dl_{j}\,e^{i\beta k\left( l_{j}\right) }  \label{charlong7}
\end{eqnarray}

Expanding the function $e^{i\beta k\left( l\right) }$ in a power series at $%
N>>1$, and taking into account the condition $\beta N=1$, we may
write
\[
\frac{1}{P}\oint dle^{i\beta k\left( l\right) }=\exp \left( i\frac{\beta }{P}%
\oint dl\,k\left( l\right) \right)
\]
that, after substitution into Eq.~(\ref{charlong7}), gives
\begin{equation}
H\left[ k\left( l\right) \right] =\exp \left( i\frac{\beta
N}{P}\oint dl\,k\left( l\right) \right)  \label{charend9}
\end{equation}

The found characteristic functional of Eq.~(\ref{charend9}) has
the form of Eq.~(\ref{charfin6}) for the function of the uniform
distribution of entry points
\begin{equation}
\Psi \left( l\right) =\frac{1}{P}  \label{uni10}
\end{equation}
This means that all the clusters have the same uniform
distribution of the entry points of Eq.~(\ref{uni10}), for which
the probability of finding a weak link at any point of the
perimeter is independent of its position.

Let us suppose that concentration of entry points into weak links
per unit perimeter length $n=\overline{N}/P$ is constant for all
clusters, and all the clusters are statistically self-similar. In
this case the mean number of entry points $\overline{N}$ along the
cluster perimeter is proportional to its length:
\begin{equation}
\overline{N}=\oint n\left( l\right) dl=nP  \label{entry11}
\end{equation}

Next step will consist in finding the relationship between the
size of a cluster and its critical current. The pinning force
corresponds to such a current at which the vortices start to break
away from the cluster. As the transport current is increasing, the
Lorentz force, which expels the magnetic flux, increases as well.
The vortices start to leave the normal phase cluster when the
Lorentz force becomes greater than the pinning force. At the same
time, growing in current will result in re-distribution of the
magnetic flux, which will penetrate deeper and deeper into a
transition layer on that side of the surrounding superconducting
space where the Lorentz force is directed (see
Fig.~\ref{fig2}(b)). In order to leave the normal phase cluster,
vortices have to reach the entry points into weak links. The exit
of the magnetic flux can be considered as the result of random
walks of vortices driven by the Lorentz force, which is pushing
them into weak links. A similar approach has been successfully
applied in analyzing the magnetic flux penetration in SQUID
arrays. \cite{dorog} The mean number of the entry points
$\overline{N}$ available on the cluster perimeter provides the
probability measure of the number of the random walk outcomes,
which are favorable for the vortex to go out. In the case of the
uniform entry point distribution, from Eq.~(\ref{entry11}) it follows that $%
\overline{N}\propto P$, so the perimeter length also represents
the probability measure of the amount of favorable outcomes for
vortex to leave the cluster. The more entry points into weak links
are accessible for random walk vortices, the smaller is the
Lorentz force required to push the flux out. Hence, we may write
the following relationship between the critical current of the
cluster and its geometric size:
\begin{equation}
I\propto \frac{1}{\overline{N}}\propto \frac{1}{P}  \label{crit12}
\end{equation}

This expression is valid for the simplest case of uniform
distribution of entry points, which is assumed to be the same for
all clusters. Such a simplification allows us to emphasize that in
the case being considered the magnetic flux is held in the normal
phase cluster by its boundary.

Thus, to deal with the distribution function of
Eq.~(\ref{proba1}), the relation between perimeter and area of
clusters should be studied. It might be natural to suppose that
the perimeter-area relation obeys the well known geometric
formula: $P\propto \sqrt{A}$. However, it would be a very rough
approximation, because this relationship holds for Euclidean
geometric objects only. As was first found in Ref.\cite{pla}, the
normal phase clusters can have fractal boundaries, which exhibit
non-Euclidean features. The fractal nature of such clusters exerts
an appreciable effect on the dynamics of a magnetic flux in
superconductors. \cite{prb,pla2,pss}

\section{A Brief Introduction to Fractal Geometry}
The notion of a fractal as an object of fractional dimension was
first introduced by Mandelbrot \cite{mandelbrot67} in 1967 and
since then it has received a lot of applications in various
domains of sciences.
\cite{mandelbrot77,mandelbrot82,mandelbrot86,feder} This concept
is closely connected to ideas of scaling and self-similarity.
Self-similarity is invariance with respect to scaling; in other
words, an invariance relative to multiplicative changes of scale.
Whereas a usual periodicity is invariance with respect to additive
translations, self-similarity is a periodicity in a logarithmic
scale.

\begin{figure}
\includegraphics{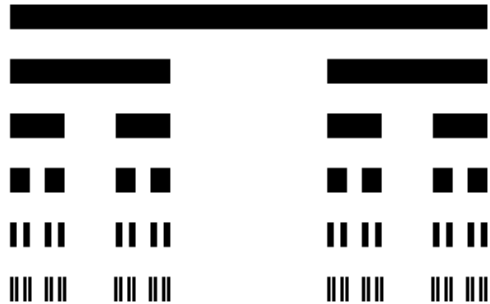}
\caption{\label{fig3} The first six steps of the Cantor dust
construction.}
\end{figure}

The simplest examples of self-similar objects are presented by so
called Cantor sets. Such a set has measure zero and, at the same
time, has so many elements that it is uncountable. The Cantor set
may be constructed in the following way. Let us draw a line
segment, and erase its middle third. Then eliminate middle third
of each remaining part, and so on. The first six steps of this
procedure are shown in Fig.~\ref{fig3}. The resulting set obtained
by endless erasing of the middle thirds of remaining intervals is
called ``Cantor dust``. \cite{mandelbrot77} The Cantor dust forms
a self-similar set: if one would take any line in Fig.~\ref{fig3},
leave out the left or right half and magnify the remainder three
times, this will result in the line segment immediately above it.
So the Cantor dust is invariant to scaling by a factor $s=3$. This
set is not only a self-similar one, but it is a fractal as well.
Let us look at that part of the set that falls within one $s$-th
part of the original set, where $s$ is a scaling factor, and ask
what fraction of the set falls into that portion? In other words,
how many subsets, each is similar to the original set, are there
if the length is subdivided into $s$ parts? The number of such
subsets is equal to $N=2$, so one-half of the original set falls
into one third of the initial length. The fractal dimension is
defined as the logarithm of the subset number divided by the
logarithm of the scaling factor: $D=\ln N/\ln s=\ln 2/\ln
3=0.631$. This formula represents the relation between the subset
number and the scaling factor: $N=s^{D}$. Thus the fractal
dimension of the Cantor dust is less than unity. This fact
reflects its dust-like consistency compared to a usual line. An
Euclidean line is such a set that if we change the length scale,
we recover the same set of points. Hence the fractal dimension of
a line coincides with the topological dimension, and both of them
are equal to unity. Generally, a fractal object has a fractional
dimension. A fractal set is such a set for which the fractal
dimension strictly exceeds its topological dimension.
\cite{mandelbrot82}

\begin{figure}
\includegraphics{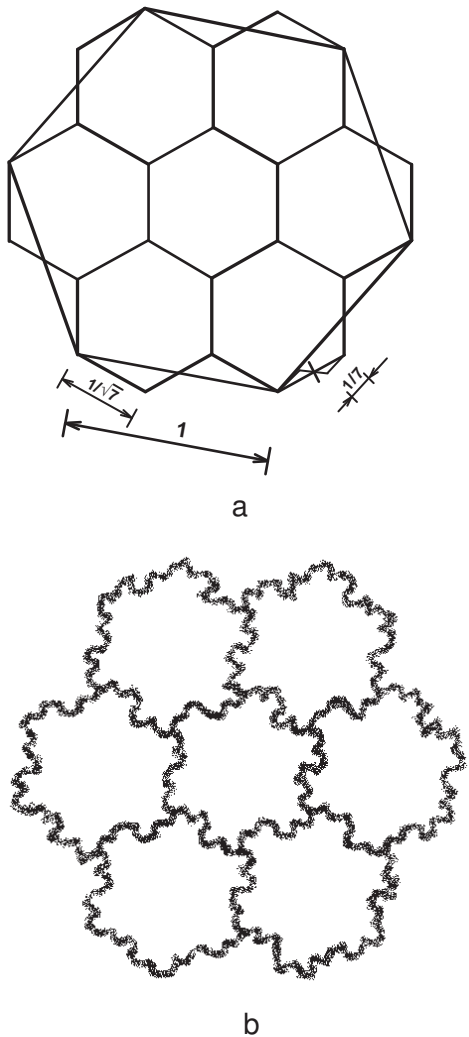}
\caption{\label{fig4} Smoother Koch island: construction (a), and
the final result (b).}
\end{figure}

Another example of the fractal set is so called smoother Koch
island. \cite{mandelbrot77} Figure~\ref{fig4}(a) shows the simple
hexagon lattice formed by a hexagon surrounded by six identical
hexagons. Let us note that each side of the big hexagon is broken
up into three straight-line segments, each of length equal to
$1/\sqrt{7}$. The total area of seven hexagons is seven times the
area of a little hexagon, whereas the perimeter of the entire
figure is three times that of a little hexagon. Note that the
total figure is not a hexagon so the figure is not similar to its
parts. Although hexagons cover the whole plane, they do not form a
hexagon again. But this peculiarity can be fixed up by making all
the figures to be similar. The first step is to make each straight
line in Fig.~\ref{fig4}(a) to be similar to one third of the
perimeter of the entire figure. For this purpose we break up each
side of a little hexagon into three segments again, each of the
length of $1/7$ of the side of the big hexagon. By repeating this
breaking process endlessly, we will arrive at a figure that is
sketched in Fig.~\ref{fig4}(b). The entire figure is similar to
its seven parts. The obtained figure is a fractal called
``smoother Koch island``. Its scaling factor is equal to
$s=\sqrt{7}$ by our construction, and the corresponding number of
subsets is equal to $N=3$, so its fractal dimension is $D=\ln
3/\ln \sqrt{7}=1.129$. The non-integer magnitude of the obtained
value is consistent with the generalized Euclid theorem about
measures of similar figures, which states that the ratios of
corresponding measures are equal when reduced to the same
dimension. \cite{mandelbrot82} Therefore, it follows that
\begin{equation}
P^{1/D}\propto A^{1/2}  \label{scaling13}
\end{equation}
which is valid both for Euclidean figures and for the fractal
ones.

Both the Cantor dust and the smoother Koch island are examples of
determinate fractals: they are self-similar by deterministic
construction. But the same approach can be applied to the
stochastic fractals, which are statistically self-similar.
\cite{mandelbrot77} The fundamental feature of any fractals, both
determinate and the stochastic ones, is that its characteristic
measures obey a scaling law that includes an exponent called the
fractal dimension. In the following context, this is the scaling
relation of Eq.~(\ref{scaling13}) where the perimeter and the
enclosed area are involved, and the fractal dimension
characterizes the cluster perimeter (so-called coastline
dimension. \cite{mandelbrot77})

\section{Fractal Geometry of Normal Phase Clusters}
Using the relation of Eq.~(\ref{scaling13}) between the fractal
perimeter and the area of the cluster, as well as formula of
Eq.~(\ref{crit12}), we can get the following relation between the
critical current of the cluster and its geometric size: $I=\alpha
A^{-D/2}$, where $\alpha $ is the form factor, and $D$ is the
fractal dimension of the cluster perimeter. In the general way the
cluster area distribution can be described by gamma distribution,
\cite{pla2,pss} which has the following cumulative probability
function:
\begin{equation}
W\left( A\right) =\left( \Gamma \left( g+1\right) \right)
^{-1}\gamma \left( g+1,\frac{A}{A_{0}}\right)  \label{cum14}
\end{equation}
where $\Gamma \left( \nu \right) $ is Euler gamma function,
$\gamma \left( \nu \right) $ is the incomplete gamma function,
$A_{0}$ and $g$ are the
parameters of gamma distribution that control the mean area of the cluster $%
\overline{A}=\left( g+1\right) A_{0}$ and its variance $\sigma
_{A}^{2}=\left( g+1\right) A_{0}^{2}$.

The case of gamma distribution of the cluster areas is of most
interest for optimizing the cluster structure of the composite
superconductors, because this distribution has two independent
parameters that can be varied in the cause of the film growth. One
of the aim of such an optimization is to get the highest
current-carrying capability of a superconductor. \cite{tpl2,pla3}

In accordance with starting formulas of Eq.~(\ref{proba1}) and
Eq.~(\ref {probi2}), gamma distribution of cluster areas of
Eq.~(\ref{cum14}) gives rise to the critical current distribution
of the form:
\begin{equation}
F\left( i\right) =\left( \Gamma \left( g+1\right) \right)
^{-1}\Gamma \left( g+1,Gi^{-2/D}\right)  \label{gammacrit15}
\end{equation}
where
\[
G\equiv \left( \frac{\theta ^{\theta }}{\theta ^{g+1}-\left(
D/2\right) \exp \left( \theta \right) \Gamma (g+1,\theta )}\right)
^{\frac{2}{D}}
\]
$\theta \equiv g+1+D/2$, $\Gamma (\nu ,z)$ is the complementary
incomplete gamma function, $i\equiv I/I_{c}$ is the dimensionless
electric current, $I_{c}=\alpha \left( A_{0}G\right) ^{-D/2}$ is
this the critical current of the transition into a resistive
state. The found cumulative probability function of Eq.~(\ref
{gammacrit15}) allows us to derive the probability density
$f\left( i\right) \equiv dF/di$ for the critical current
distribution:
\begin{equation}
f(i)=\frac{2G^{g+1}}{D\Gamma (g+1)}i^{-\left( 2/D\right)
(g+1)-1}\exp \left( -Gi^{-2/D}\right)  \label{dens16}
\end{equation}
This distribution allows us to fully describe the effect of the
transport current on the trapped magnetic flux taking into account
the fractal properties of the normal phase clusters.

\begin{figure}
\includegraphics[height=3in]{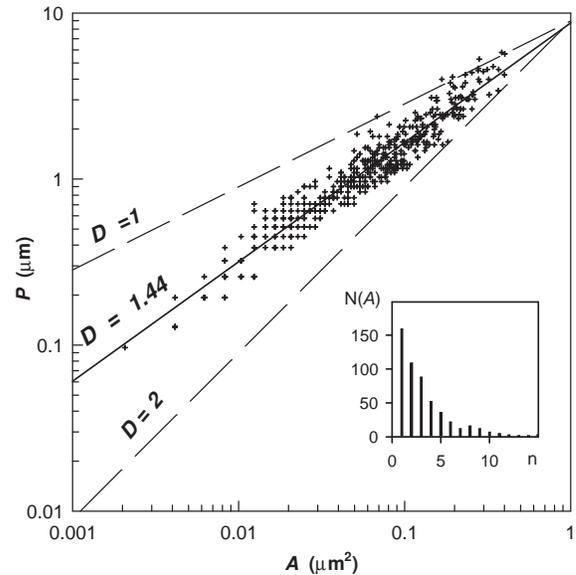}
\caption{\label{fig5} Perimeter-area relationship for the normal
phase clusters with fractal boundary. The solid line indicates the
least square regression line, two dotted lines display the range
of slope that the perimeter-area curves can have for any possible
fractal dimension $D$. ($D=1$ - for clusters of Euclidean
boundary, $D=2$ - for clusters of boundary with the maximum
fractality). The inset represents the histogram of the cluster
area sampling with rank spacing of 28680\thinspace nm$^{2}$.}
\end{figure}

\begin{table*}
\begin{center}
  \begin{tabular}{ccc}
    \hline\hline
    Mean $A$, nm$^{2}$ & $76540$ \\
    Sample standard deviation of $A$, nm$^{2}$ & $72620$ \\
    Min value of $A$, nm$^{2}$ & $2069$ \\
    Max value of $A$, nm$^{2}$ & $401500$ \\ \hline
    Mean $P,$nm & $1293$ \\
    Sample standard deviation of $P$, nm & $962$ \\
    Min value of $P$, nm & $96$ \\
    Max value of $P$, nm & $5791$ \\ \hline Correlation coefficient &
    $0.929$ \\ \hline Estimated fractal dimension $D$ & $1.44$ \\
    \hline Standard deviation of $D$ & $0.02$ \\ \hline\hline
  \end{tabular}
\end{center}
\caption{Statistics of normal phase clusters\label{tabl1}}
\end{table*}
In order to clear up how the developed approach can be used in
practice, the geometric probability analysis of electron
photomicrographs of superconducting films was carried out. For
this purpose electron photomicrographs of YBCO film prepared by
magnetron sputtering have been scanned. The normal phase has
occupied 20\% of the total surface only, so the transport current
can flow through the sufficiently dense percolation
superconducting cluster. The perimeters and areas of clusters have
been measured by covering their digitized pictures with a square
grid of spacing $60\times 60\,$nm$^{2}$. The results of the
statistical treatment of these data are presented in
Table~\ref{tabl1} as well as in Fig.~\ref{fig5}. The primary
sampling has contained 528 normal phase clusters located on the
scanned region of a total area of 200~$\mu $m$^{2}$. The
distribution of the cluster areas is fitted well to exponential
cumulative probability function, as is shown on the histogram in
the inset of Fig.~\ref{fig5}. The number of clusters that fall
within the assigned rank is plotted on the ordinate of this graph;
the rank number is plotted on the abscissa. A high skewness
(1.765) as well as the statistically insignificant (5\%)
difference between the sample mean area of the cluster and the
standard deviation also attests that there an exponential
distribution of the cluster areas holds true. This distribution is
a special case of gamma distribution for which $g=0$, so the
cumulative probability function of Eq.~(\ref{cum14}) can be
simplified to the following form:
\begin{equation}
W(A)=1-\exp \left( -\frac{A}{\overline{A}}\right)  \label{exp17}
\end{equation}
The exponential distribution has only one characteristic parameter
- the mean cluster area $\overline{A}$. The obtained data allow us
to find the perimeter-area relation for the normal phase clusters.
All the points in Fig.~\ref{fig5} fall on a straight line in
double logarithmic scale with correlation coefficient of 0.929.
Accordingly to the scaling relation of Eq.~(\ref{scaling13}), the
slope of the regression line gives the estimate of fractal
dimension of the cluster perimeter, which is equal to $D=1.44\pm
0.02$. The graph in Fig.~\ref{fig5} shows that the scaling law for
perimeter and area, which is inherent to fractals, is valid in the
range of almost three orders of magnitude in cluster area. This
relation between perimeter and area is consistent with the
generalized Euclid theorem, so we can see that the ratios of
perimeters and areas are equal when reduced to the same dimension.
The scaling perimeter-area behavior means that there is no
characteristic length scale in the range of two orders of linear
size of the normal phase cluster. Whatever the shape and size of
the clusters may be, all the points fall closely on the same
straight line in logarithmic scale; so that there are no apparent
kinks or bends on the graph.

This point that the found value of the fractal dimension differs
appreciably from unity engages a great attention. What this means
is that the fractal properties of the cluster boundary are of
prime importance here. Two dotted lines in Fig.~\ref{fig5} bound
the range of the slopes that the dependencies of the perimeter on
the cluster area can have for any arbitrary fractal dimension. The
least slope (upper line) corresponds to Euclidean clusters, which
have the fractal dimension equal to topological one ($D=1$), the
most one (lower line) relates to the clusters of the greatest
possible fractal
dimension, which is equal to the topological dimension of a smooth surface ($%
D=2$). Such a fractal dimension is inherent, for example, in Peano
curves, which fill the whole plane. \cite{mandelbrot77} Whatever
the geometric morphological properties of clusters may be, the
slope of their perimeter-area graphs will be always bounded by
these two limiting lines. The found dependence ($D=1.44$) runs
just between them.

\begin{figure}
\includegraphics[height=3in]{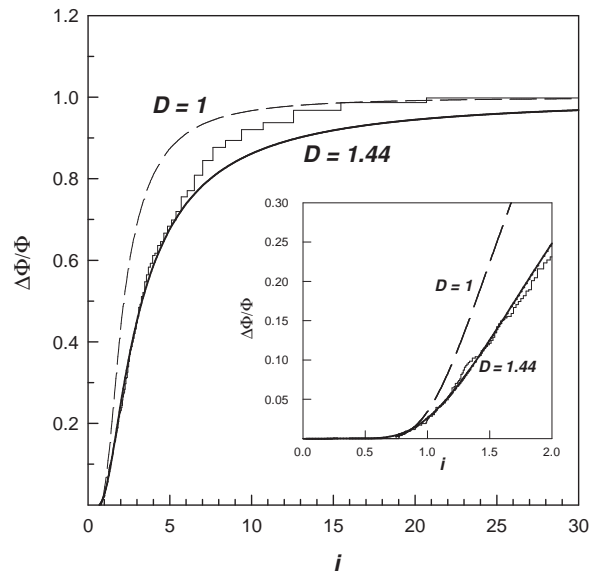}
\caption{\label{fig6} Effect of a transport current on the
magnetic flux trapped in fractal clusters of a normal phase. The
solid line shows the decrease in trapped flux for the fractal
clusters of coastline dimension $D=1.44$; the dotted line
corresponds to the case of Euclidean clusters ($D=1$); step line
is the sample empirical function of critical current
distribution.}
\end{figure}

The geometric probability properties of the normal phase clusters
are responsible for main features of the critical current
statistical distribution. Now, knowing the fractal dimension of
the cluster boundaries, the change in the trapped magnetic flux
caused by the transport current can be found with aid of
Eq.~(\ref{probi2}). The exponential distribution of the cluster
areas of Eq.~(\ref{exp17}) gives rise to the
exponential-hyperbolic distribution of critical currents
\begin{equation}
F(i)=\exp \left( -\left( \frac{2+D}{2}\right)
^{2/D+1}i^{-2/D}\right) \label{exhyp18}
\end{equation}
which follows from Eq.~(\ref{gammacrit15}) at $g=0$. Now the
critical current of the resistive transition $I_{c}$, which
appears in the expression for the dimensionless electric current
$i\equiv I/I_{c}$, can be found from a simpler formula:
$I_{c}=\left( 2/\left( 2+D\right) \right) ^{\left( 2+D\right)
/2}\alpha \left( \overline{A}\right) ^{-D/2}$. The effect of a
transport current on the trapped magnetic flux is illustrated in
Fig.~\ref{fig6} for the case of Euclidean clusters (dotted curve)
as well as for the clusters of found fractal dimension $D=1.44$
(solid curve).

In order to get the relationship between the dynamics of the
trapped magnetic flux and the geometric morphological properties
of the superconducting structure the empirical function of the
distribution of the critical currents $F^{\ast }=F^{\ast }(i)$ has
been found. First, the empirical distribution function $W^{\ast
}=W^{\ast }(A)$ for the sampling of
the areas of the normal phase clusters has been obtained. The value of $%
W^{\ast }(A)$ was calculated for each order statistic as the
relative number of clusters of area smaller than a given value
$A$. Next, the empirical distribution of the critical currents was
computed for the same order statistics using the formulas:

\[
\left.
\begin{array}{c}
F^{\ast }=1-W^{\ast } \\
i=\left( \frac{2+D}{2}\right) ^{\left( 2+D\right) /2}\left( \frac{\overline{A%
}}{A}\right) ^{D/2}
\end{array}
\right\}
\]

This function, shown in Fig.~\ref{fig6} by the step line, gives a
statistical estimate of the cumulative probability function of
Eq.~(\ref {exhyp18}). As is seen from this figure, both
distributions coincide well in the range of currents $i<6$.
Starting with the value of the current $i=6$ the crossover from
the fractal regime to the Euclidean one is observed. This
transition into the Euclidean regime is over at large transport
currents, when the magnetic flux changes mainly for the breaking
of the vortices away from the small clusters (as the smaller
clusters have the larger pinning force). The observed crossover
has its origin in the finite resolution capability of measuring
the cluster sizes. When estimating the fractal dimension, we have
to take into account that the resolution of the measurement of any
geometric sizes is finite. The peculiarity of the topologically
one-dimensional fractal curve is that its measured length depends
on the measurement accuracy. \cite{mandelbrot77} In our case such
a fractal curve is represented by the boundary of the normal phase
cluster. That is why just the statistical distribution of the
cluster areas, rather than their perimeters, is fundamental for
finding the critical current distribution. The topological
dimension of perimeter is equal to unity and does not coincide
with its fractal dimension, which strictly exceeds the unity.
Therefore the perimeter length of a fractal cluster is not well
defined, because its value diverges as the yardstick size is
reduced infinitely. On the other hand, the topological dimension
of the cluster area is the same as the fractal one (both are equal
to two). Thus, the area restricted by the fractal curve is a
well-defined finite quantity.

Taking into account the finite resolution effect, the
perimeter-area relationship of Eq.~(\ref{scaling13}) can be
re-written in the following form
\begin{equation}
P(\delta )\propto \delta ^{1-D}\left( A\left( \delta \right)
\right) ^{D/2} \label{rescaling19}
\end{equation}
where $\delta $ is the yardstick size used to measure this length.
This relation holds true when the yardstick length is small enough
to measure accurately the boundary of all smallest clusters in
sampling. When the resolution is deficient, the Euclidean part of
the perimeter length will dominate the fractal one, so there is no
way to find the fractal dimension using the scaling relation of
Eq.~(\ref{rescaling19}). It means that if the length of a fractal
curve was measured too roughly with the very large yardstick, its
fractal properties could not be detected, and therefore such a
geometric object would be manifested itself as Euclidean one. It
is just the resolution deficiency occurs at the crossover point in
Fig.~\ref{fig6}. Starting with the cluster area less than some
value (which corresponds to the currents of $i>6$) it is
impossible to measure all `` skerries'' and `` fjords'' on the
cluster coastlines, whereas all the clusters of area less than the
size of the measuring cell (3600$\,$nm$^{2}$ that relates to the
currents of $i>23$), exhibit themselves as objects of Euclidean
boundaries. This resolution deficiency can be also observed in
Fig.~\ref{fig5}: some points at its lower left corner are arranged
discretely with the spacing equal to the limit of resolution
(60$\,$nm), because some marks for smallest clusters coincide for
the finite resolution of the picture digitization procedure.

The fractal dimension was found above by means of regression
analysis of the whole primary sampling, where the very small
clusters of sizes lying at the breaking point of the resolution
limit were also included. In order to evaluate how the finite
resolution affects the accuracy of the estimation of the fractal
dimension, all the points, for which the resolution deficiency was
observed, were eliminated from the primary sampling. So the
truncated sampling has been formed in such a way that only 380
clusters, for which the resolution deficiency is not appeared yet,
have been selected from the primary sampling. The least squares
treatment of these perimeter-area data gives the adjusted
magnitude of the fractal dimension: $D=1.47\pm 0.03$. The found
value virtually does not differ from the previous one within the
accuracy of the statistical estimation, whereas the correlation
coefficient (which becomes equal to 0.869) falls by six hundredth
only. Therefore, we can conclude that the found estimate of the
fractal dimension is robust.

It is worthy of note that the above-described resolution
deficiency refers only to the procedure of the cluster geometric
sizes measurement. As may be seen from the Table~\ref{tabl1}, the
characteristic sizes of the normal phase clusters far exceed both
the coherence length and the penetration depth. Therefore, any
effects of finite resolution related to the size of the vortices
that scan the boundaries of the clusters in searching for weak
links may be neglected.

\section{Fractal Superconducting Structures in a Resistive State}
\begin{figure}
\includegraphics[height=2.25in]{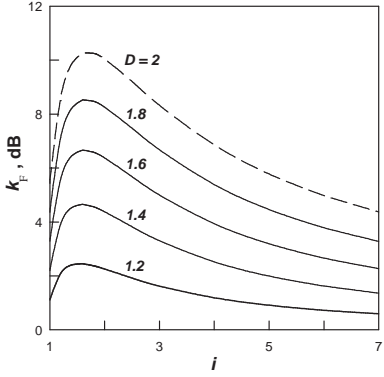}
\caption{\label{fig7} Pinning gain for the different fractal
dimension of the cluster perimeter in the case of
exponential-hyperbolic distribution of the critical currents
($g=0$).}
\end{figure}

So, it has been revealed that the fractality of the cluster
boundaries intensifies the pinning. As can be seen from
Fig.~\ref{fig6}, the decrease in the trapped magnetic flux at the
same value of the transport current is less for larger fractal
dimension. The pinning amplification can be characterized by the
pinning gain factor
\begin{equation}
k_{\Phi }\equiv 20\log \frac{\Delta \Phi \left( D=1\right)
}{\Delta \Phi \left( D\right) }  \label{gain20}
\end{equation}
which is equal to relative decrease (in decibels) in the fraction
of magnetic flux broken away from fractal clusters of fractal
dimension $D$ compared to the case of Euclidean ones. The
dependencies of the pinning gain on the transport
current for different fractal dimension at $g=0$ are shown in Fig.~\ref{fig7}%
. The highest amplification (about 10\thinspace dB) is reached
when the cluster boundaries have the greatest possible fractality.
Figure~\ref{fig7} demonstrates that with increase in fractal
dimension the trapped magnetic flux is changed less and less by
the action of the transport current. The pinning gain of
Eq.~(\ref{gain20}) characterizes the properties of a
superconductor in the range of the transport currents
corresponding to a resistive state ($i>1$). At smaller current the
total trapped flux remains unchanged (see Fig.~\ref{fig6}) for
lack of pinning centers of such small critical currents, so the
breaking of the vortices away has not started yet. When the
vortices start to leave the normal phase clusters and move through
the weak links, their motion induces an electric field, which, in
turn, creates the voltage drop across the sample. Therefore, the
passage of electric current is accompanied by the energy
dissipation. As for any hard superconductor (that is to say,
type-II, with pinning centers) this dissipation does not mean the
destruction of phase coherence yet. Some dissipation always
accompanies any motion of a magnetic flux that can happen in a
hard superconductor even at low transport current. Therefore the
critical current in such materials cannot be specified as the
greatest non-dissipative current. The superconducting state
collapses only when a growth of dissipation becomes avalanche-like
as a result of thermo-magnetic instability.

By virtue of the fact that any motion of the magnetic flux causes
the energy dissipation in superconductors, the question of how
such a process could be prevented, or only suppressed, is of prime
practical importance. The study of resistive state peculiarities
leads to the conclusion that the cluster fractality exerts
influence on the electric field induced by the flux motion. The
found distribution of the critical currents allows us to find the
electric field arising from the magnetic flux motion after the
vortices have been broken away from the pinning centers. Inasmuch
as each normal phase cluster contributes to the total critical
current distribution, the voltage across a superconductor can be
represented as the response to the sum of effects made by the
contribution from each cluster. Such a response can be expressed
as a convolution integral
\begin{equation}
\frac{V}{R_{f}}=\int\limits_{0}^{i}\left( i-i^{\prime }\right)
f\left( i^{\prime }\right) di^{\prime }  \label{volt21}
\end{equation}
where $R_{f}$ is the flux flow resistance. The similar approach is
used universally in all the cases where the distribution of the
depinning currents takes place.
\cite{warnes,wordenweber,sst,brown} The subsequent consideration
will be essentially concentrated on the consequences of the
fractal nature of the normal phase clusters specified by the
distribution of Eqs.~(\ref{gammacrit15}), (\ref{exhyp18}), so all
the problems related to possible dependence of the flux flow
resistance $R_{f}$ on a transport current will not be taken up
here.

After substitution of the critical current distribution of Eq.~(\ref{dens16}%
) into the expression of Eq.~(\ref{volt21}), upon integration, we
get the final expression for the voltage-current ({\it V-I})
characteristics in the general case of the gamma-distribution of
the cluster areas:
\begin{eqnarray}
\frac{V}{R_{f}}&=&\frac{1}{\Gamma \left( g+1\right) }\Biggl(
i\Gamma \left( g+1,Gi^{-2/D}\right)  \nonumber \\
&&-G^{D/2}\Gamma \left( g+1-\frac{D}{2},Gi^{-2/D}\right) \Biggr)
\label{gammavi22}
\end{eqnarray}

In the special case of exponential cluster area distribution
($g=0$) the general formula of Eq.~(\ref{gammavi22}) can be
simplified:
\begin{eqnarray}
\frac{V}{R_{f}}&=&i\exp \left( -C\,i^{-2/D}\right)  \nonumber \\
&&-C^{D/2}\Gamma \left( 1-\frac{D}{2},C\,i^{-2/D}\right)
\label{expvi23}
\end{eqnarray}
where $C\equiv \left( \left( 2+D\right) /2\right) ^{2/D+1}$.

In extreme cases of Euclidean clusters and clusters of the highest
fractality the expression of Eq.~(\ref{expvi23}) can be further
transformed:

(i) Clusters of Euclidean boundary ($D=1$ at $g=0$):
 \[
\frac{V}{R_{f}}=i\exp \left( -\frac{3.375}{i^{2}}\right) -\sqrt{3.375\pi }%
\mathrm{erfc}\left( \frac{\sqrt{3.375}}{i}\right)
\]

where $\mathrm{erfc}\left( z\right) $ is the complementary error function.

(ii) Clusters of boundary with the maximum fractality ($D=2$ at
$g=0$):
\[
\frac{V}{R_{f}}=i\exp \left( -\frac{4}{i}\right) +4%
\mathop{\rm Ei}%
\left( -\frac{4}{i}\right)
\]

where $%
\mathop{\rm Ei}%
\left( z\right) $ is the exponential integral function.

\begin{figure}
\includegraphics[height=2.25in]{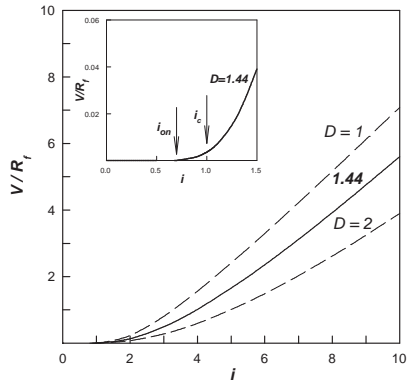}
\caption{\label{fig8} Voltage-current characteristics of
superconductors with fractal clusters of a normal phase for
exponential-hyperbolic distribution of the critical currents
($g=0$). Dotted lines correspond to extreme cases of Euclidean
clusters ($D=1$) and clusters of the most fractality ($D=2$);
solid line -
to superconductor containing the normal phase clusters of fractal dimension $%
D=1.44$. The inset shows the region near the resistive transition
in an enlarged scale. The initial dissipative range between
$i_{on}$ and $i_{c}$ is clearly seen.}
\end{figure}

The {\it V-I} characteristics calculated using Eq.~(\ref
{expvi23}) for several values of $D$ are presented in
Fig.~\ref{fig8}. Two dotted lines bound the region the {\it V-I}
characteristics can fall within for any possible values of fractal
dimension. The solid curve represent the {\it V-I} characteristic
of a superconductor with fractal clusters of previously obtained
fractal dimension $D=1.44$. The inset in Fig.~\ref{fig8} shows the
region of resistive transition under magnification. As may be seen
from this graph, the critical current $i_{c}$ is preceded by some
initial region of the finite voltage drop starting with $i_{on}$,
so the resistive transition
is not absolutely abrupt. The existence of this initial section on the {\it %
V-I} characteristic arises from the peculiarities of the fractal
distribution in the range of small currents. It has been just a
similar initial region of fractal dissipation has been observed in
high-resolution measurements of dynamical resistance of HTS-normal
metal composites (BPSCCO-Ag) as well as in polycrystalline YBCO
and GdBCO samples. \cite{prester,prester2}

\begin{figure}
\includegraphics[height=2.25in]{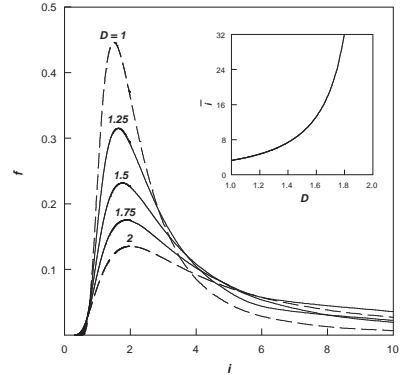}
\caption{\label{fig9} Influence of the fractal dimension of the
cluster perimeter on the exponential-hyperbolic distribution of
the critical currents ($g=0$). The inset shows the dependence of
the mean critical current $\overline{i}$ on the fractal
dimension.}
\end{figure}

All the curves of {\it V-I} characteristics are virtually starting
with the transport current value equal to unity. When the current
increases the trapped flux remains unchanged until the vortices
start to break away from the pinning centers. As long as the
magnetic flux does not move, no electric field is arisen.
Figure~\ref{fig8} shows that the fractality reduces appreciably an
electric field arising from the magnetic flux motion. This effect
is especially strong in this range of the currents ($1<i<3$),
where the pinning enhancement also has a maximum (see also
Fig.~\ref{fig7}). Both these effects have the same nature,
inasmuch as their reason lies in the peculiarities of the critical
current distribution. The influence of the fractal dimension of
the cluster boundary on the critical current distribution is
demonstrated in Fig.~\ref{fig9}. As may be seen from this graph,
the bell-shaped curve of the distribution broadens out, moving
towards greater magnitudes of current as the fractal dimension
increases. It means that more and more of the small clusters,
which can best trap the magnetic flux, are being involved in the
game. Hence the number of vortices broken away from pinning
centers by the Lorentz force is reducing, so the smaller part of a
magnetic flux can flow. The smaller part of a magnetic flux can
flow, the smaller electric field is created. In turn, the smaller
the electric field is, the smaller is the energy dissipated when
the transport current passes through the sample. Therefore, the
decrease in heat-evolution, which could cause transition of a
superconductor into a normal state, means that the
current-carrying capability of the superconductor containing such
fractal clusters is enhanced.

The re-distribution of probability with increasing in the fractal
dimension can be characterized by the statistical moments of the
critical current distribution, namely, the mean critical current
\[
\overline{i}=G^{D/2}\frac{\Gamma \left( g+1-D/2\right) }{\Gamma
\left( g+1\right) }
\]
and the variance of critical currents
\[
\sigma _{i}^{2}=G^{D}\left( \frac{\Gamma \left( g+1-D\right)
}{\Gamma \left( g+1\right) }-\left( \frac{\Gamma \left(
g+1-D/2\right) }{\Gamma \left( g+1\right) }\right) ^{2}\right)
\]
As may be seen from the inset in Fig.~\ref{fig9}, the mean
critical current grows nonlinearly with increase in the fractal
dimension that reflects the broadening of the ``tail`` of the
critical current distribution. This ``tail`` is so elongated that
the variance of critical
currents becomes infinite in the range of fractal dimensions $D\ge g+1$%
. At the same time, the mode of the critical current distribution
is a finite function, which increases monotonically in the
range of all possible values of fractal dimension: {\it %
mode\thinspace }$f\left( i\right) =\left( G/\theta \right)
^{D/2}$. The distributions with divergent variance are known in
probability theory - the classic example of that kind is Cauchy
distribution. \cite{hudson} However, such an anomalous feature of
the critical current distribution of Eq.~(\ref{dens16}) is of
special interest, inasmuch as the current-carrying capability of a
superconductor would be expected to increase just in the region of
giant dispersion. Then the statistical distribution of critical
currents has a very broad ``tail`` containing the contributions
from the clusters of the highest depinning currents.

\section{Conclusion}
So, in the present work the fractal nature of the normal phase
clusters is revealed, and relation between the fractal properties
of the clusters and dynamics of the trapped magnetic flux is
established. The fractal distribution of the critical current is
obtained. It is found that the fractality of cluster boundary
strengthens the flux pinning and thereby
hinders the destruction of superconductivity by the transport current. {\it %
V-I} characteristics of fractal superconducting structures in a
resistive state are obtained. It is revealed that the fractality
of the boundaries of the normal phase clusters reduces the
electric field arising from magnetic flux motion, and thereby
raises the critical current of a superconductor.

At the same time, the following problems are open for further
development in this field. First, a central point of the presented
approach is that the magnetic flux is held on the boundaries of
clusters. In experiments with polycrystalline magnesium diboride
\cite{larbalestier} it has been found that in this new
superconductor the similar situation takes place, but instead of
the cluster boundaries, the grain boundaries act in the same way.
Second, the fractal distribution of the critical currents have
anomalous statistical properties caused by a divergence of some
its statistical moments so this point is worthy for further
investigations. Then, the special cases of the distribution of
entry points into weak links, which can be realized in anisotropic
composites, are of interest. First of all, we mean the entry point
distributions that take place in superconducting tapes and wires.
Last, obtained results open the possibilities for increasing the
critical current value of percolative superconductors by
optimizing their geometric morphological properties, and enable to
improve the technology of preparation of superconducting films of
high critical current value. The study of these questions will
allow us to get a better insight into the vortex dynamics in the
superconductors containing fractal clusters.

\begin{acknowledgments}
This work is supported by the Saint Petersburg Scientific Center
of the Russian Academy of Sciences.
\end{acknowledgments}

\bibliography{ARW1}

\end{document}